\documentclass[twocolumn,aps,prb,groupedaddress]{revtex4-2}

\usepackage{graphicx}
\usepackage{color}
\usepackage{amsmath}
\usepackage{enumitem}
\usepackage{amssymb}
\usepackage{hyperref}
\usepackage{cancel}
\usepackage{ulem}
\usepackage{multirow}

\newcommand{\be}{\begin{equation}}
	\newcommand{\ee}{\end{equation}}

\newcommand{\bea}{\begin{eqnarray}}
	\newcommand{\eea}{\end{eqnarray}}

\newcommand{\p}{\partial}

\newcommand{\la}{\left\langle}
\newcommand{\ra}{\right\rangle}
\newcommand{\lb}{\left[}
\newcommand{\rb}{\right]}
\newcommand{\lp}{\left(}
\newcommand{\rp}{\right)}

\renewcommand{\vec}[1]{{\boldsymbol #1}}

\begin{document}

\title{Collinear scattering and long-lived excitations in two-dimensional electron fluids}
%
\author{Serhii Kryhin and Leonid Levitov}
\affiliation{Department of Physics, Massachusetts Institute of Technology, Cambridge, MA 02139}

\begin{abstract}
For a long time, it has been thought that 2D Fermi gases could support long-lived excitations, thanks to the collinear quasiparticle scattering controlled by phase space constraints at a 2D Fermi surface. We present a direct calculation that reveals such excitations. The excitation lifetimes are found to exceed the fundamental bound set by Landau Fermi-liquid theory by a factor as large as $(T_F/T)^\alpha$ with $\alpha \approx 2$. These excitations represent Fermi-surface modulations of an odd parity, one per each odd angular momentum. To explain this surprising behavior, we employ a connection between the linearized quantum kinetic equation and the dynamics of a fictitious quantum particle moving in a 1D reflectionless ${\rm sech^2}$ potential. In this framework, we identify the long-lived excitations in Fermi gases as zero modes that arise from supersymmetry. 
\end{abstract}

\date{\today}

\maketitle

Microscopic theory of carrier collisions in two-dimensional (2D) electron systems 
is essential for the field of electron hydrodynamics, an area that has made significant progress in recent years \cite{Guerrero-Becerra2019,Hasdeo2021,Muller2009,Principi2016,Scaffidi2017,Narozhny2019, Alekseev2020,Toshio2020,Narozhny2021,Tomadin2014,Principi2016,Lucas2018, Qi2021,Cook2021,Valentinis2021a,Valentinis2021b,DasSarma2022,HGuo2017,AShytov2018,Nazaryan2021}.   
Theory of Fermi liquids that links carrier collision rates and quasiparticle lifetimes is generally considered to be comprehensive and complete. However, recent research has challenged the widely-held belief that the theory is entirely free of gaps and inconsistencies
\cite{Gurzhi1995,Buhmann2002,Ledwith2017,tomogrph,Ledwith2019}. 
Specifically, this literature indicates that Landau's $T^2$ scaling law, which describes  quasiparticle decay in three-dimensional Fermi-liquids at low temperatures, may not hold true for 2D metals. This happens because 2D fermions display two-body scattering of a unique collinear character, arising due to kinematic phase space constraints at the Fermi surface. These findings have interesting implications for our understanding of Fermi-liquids, as they suggest that the behavior of quasiparticles in 2D materials may differ significantly from that in 3D materials. Quenching of Landau's $T^2$ damping for certain excitations points to new ways for extending coherence in electron systems. The aim of this work is to validate these predictions through a direct calculation. 

The collinear behavior in 2D raises an interesting comparison with one-dimensional (1D) systems, where collinear scattering causes quasiparticles to have a short lifespan. 
Interactions in 1D systems destroy the Fermi-liquid state, leading to a state known as the Tomonaga-Luttinger state \cite{Giuliani_book,Giamarchi_book}.
The collinear processes in 2D metals 
take on a role which is a complete opposite of that in 1D liquids. 
These processes give a giant boost to quasiparticle lifetimes and can be said to produce a ``super-Fermi-liquid'' that harbors 
a unique family of excitations with exceptionally long lifetimes, 
exceeding by orders of magnitude those familiar from Fermi-liquid theory. 
The unique 
behavior arising from these processes endows the kinetics of 2D fermions with angular memory and gives rise to peculiar `tomographic' response effects\cite{Ledwith2017,tomogrph,Ledwith2019}.  

The presence of long-lived degrees of freedom can significantly enhance the response to weak perturbations, leading to the emergence of long-lasting collective memory effects and novel hydrodynamic modes. In this regard, recent work \cite{Kryhin2023} predicts the existence of a distinct family of viscous modes characterized by non-Newtonian viscosity. These modes were not anticipated by earlier studies, highlighting the importance of the long-lived degrees of freedom originating from collinear scattering for 2D electron transport.

\begin{figure}[t]
\centering
\includegraphics[width=0.95\columnwidth]{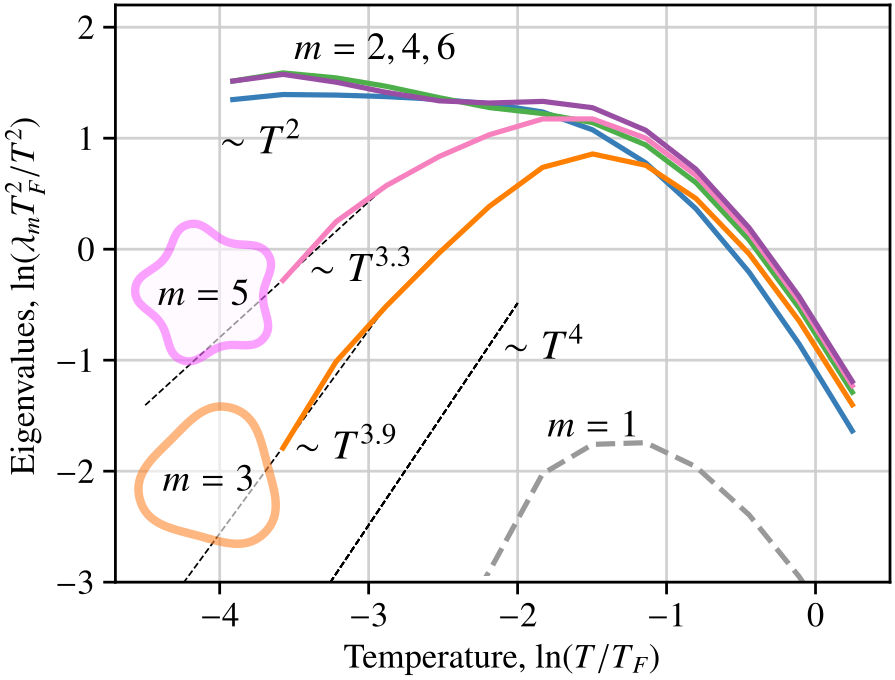} 
\caption{Decay rates for different angular harmonics of particle distribution, scaled by $T^2$, 
vs. temperature. 
Shown are dimensionless eigenvalues $\lambda_m$ related to the decay rates through $\gamma_m=A p_F^2 \lambda_m $, see Eq.\eqref{eq:int_for_calc}.
Double-log scale is used to facilitate comparison of 
disparate time scales. 
Decay rates for even-$m$ harmonics obey 
a $T^2$ scaling at $T\ll T_F$. 
Decay rates for odd-$m$ harmonics are markedly smaller than 
those for even $m$ and show ``super-Fermi-liquid'' scaling strongly deviating from $T^2$. Odd-$m$ decay rates 
can be approximated as $T^\alpha$ with $\alpha>2$. 
An even/odd asymmetry in the rates and the suppression of decays for odd $m$ 
is seen already at $T \lesssim 0.16 T_F$. 
}
\label{fig:fig1}
\end{figure}

The emergence of novel time scales is particularly evident in a system with isotropic band dispersion and a circular Fermi surface. In such a system, various excitations correspond to distinct angular harmonics of Fermi surface modulations that evolve in space and time as
\[
\delta f(\vec p,\vec x,t)\sim {\textstyle \sum_m} 
\alpha_m(\epsilon,\vec x,t) \cos m\theta +\beta_m(\epsilon,\vec x,t) \sin m\theta
,
\] 
where $\theta$ is the angle parameterizing the Fermi surface. The microscopic decay rates, illustrated in Fig.\ref{fig:fig1}, govern dynamics of spatially-uniform excitations, $\alpha_m$, $\beta_m\sim e^{-\gamma_m t}$. As evident in Fig.\ref{fig:fig1}, at low temperatures $T\ll T_F$ the lifetimes of these modes greatly exceed the ones for even $m$, showing strong departure from conventional Fermi-liquid scaling. 
The decay rates in Fig.\ref{fig:fig1} are obtained by a direct calculation that treats quasiparticle scattering 
exactly, using a method that does not rely on the small parameter $T/T_F\ll1$. The odd-$m$ decay rates display scaling $\gamma\sim T^\alpha$ with super-Fermi-liquid exponents $\alpha>2$. In our analysis we find $\alpha$ values close to $4$, i.e. the odd-$m$ rates are strongly suppressed  compared to the even-$m$ rates, $\gamma_{\rm odd}/\gamma_{\rm even}\sim (T/T_F)^2$. 

Is there a simple explanation for why the odd-m harmonics are found to be  long-lived? These harmonics are essentially the perturbations in particle momentum distribution associated with angle-resolved current, the quantities odd under $\vec p\to-\vec p$ that can take different values on different patches of the Fermi surface. The significance of these ``tomographic'' quantities is that they are approximately conserved when two-body collisions have a strongly collinear character. In comparison, for two-body collisions in a classical gas, the $p$-wave (m=1) harmonic of current is conserved, whereas higher-order harmonics ($m=3$, $5$, etc.) are non-conserved. However, in Fermi gases, as discussed below, the collisions are strongly collinear. This property endows all angular harmonics of current, that is the odd-$m$ harmonics of particle distribution, with exceptionally long lifetimes.

It is worth noting that the absence of Landau's $T^2$ damping in odd-$m$ modes may seem to contradict the results in the literature on excitation lifetimes in 2D Fermi gases, which predict that quasiparticle lifetimes are diminished by collinear scattering, as revealed by self-energy calculations of Green's functions \cite{Hodges1971,Chaplik1971,Bloom1975,Giuliani1982,Zheng1996,Menashe1996,Chubukov2003}. 
The predicted decay rates were found to be faster by a logarithmic factor $\log(T_F/T)$ compared to the conventional $T^2$ rates. Surprisingly, the self-energy approach fails to account for the existence of long-lived odd-$m$ excitations. This is unexpected because it is commonly assumed that there is a single timescale that characterizes decay for all low-energy excitations. However, as shown in Fig. \ref{fig:fig1}, the odd-$m$ and even-$m$ modes have drastically different lifetimes that exhibit different scaling behavior with respect to $T$. The conventional self-energy approach falls short in effectively addressing this particular situation as it primarily emphasizes the fastest decay pathways, thereby neglecting the presence of long-lived excitations. Surprisingly, despite an extensive and fervent interest in the field of Fermi liquids spanning over 60 years, the long-lived excitations have been overlooked in the existing literature.

We want to emphasize that the collinear processes that generate long-lived excitations are universal and largely independent of the specifics of two-body interactions or particle dispersion characteristics. 
The existence of long-lived excitations is a robust property that persists for non-circular Fermi surfaces, as long as the surface distortion is not significant. 
This is due to the presence of inversion symmetry, which separates Fermi surface modulations into even and odd parity modes. Similar to the self-energy analysis \cite{Hodges1971,Chaplik1971,Bloom1975,Giuliani1982,Zheng1996,Menashe1996,Chubukov2003}, the difference in lifetimes between these mode types is identical to that observed in circular Fermi surfaces.

We also note that in certain electron systems, collinear dynamics can accelerate quasiparticle decay by allowing particles, by traveling side by side, interact more strongly. 
This is well-documented in Dirac bands where collinear dynamics arising from linear band dispersion shortens carrier lifetimes and accelerates dynamics \cite{Gonzalez1996,Brida2013,Song2013,Li2013,Briskot2014,Trushin2016,Lewandowski2018,Kiselev2019}. In our problem, an entirely different behavior arises due to collinear scattering and phase space constraints, 
the effects that dominate at a 2D Fermi surface but are of little importance for highly excited states in Dirac bands.

The analysis presented below is based on the Fermi-liquid transport equation that accounts for the kinetics of two-body collisions constrained by fermion exclusion,
\be\label{eq:kinetic_eqn}
\frac{d f_1}{dt}+[f_1,H]=\sum_{21'2'}\lp w_{1'2'\to 12}-w_{12\to 1'2'}\rp, 
\ee
where $f(\vec p,\vec r,t)$ is fermion distribution, $[f,H]$ denotes the Poisson bracket $\nabla_{\vec r} f\nabla_{\vec p}\epsilon-\nabla_{\vec r}\epsilon \nabla_{\vec p}f$. 
The right-hand side is the rate of change of the occupancy of a state $\vec p_1$, given as a sum of the gain and loss contributions resulting from the two-body scattering processes $12\to 1'2'$ and $1'2'\to 12$.  
Fermi's golden rule yields 
\be\label{eq:Golden_Rule}
w_{1'2'\to 12}=\frac{2\pi}{\hbar}|V_{12,1'2'}|^2 
\delta_\epsilon \delta_{\vec p}(1-f_{1})(1-f_{2}) f_{1'}f_{2'},
\ee 
where the delta functions 
$\delta_\epsilon=\delta(\epsilon_1+\epsilon_2-\epsilon_{1'}-\epsilon_{2'})$, 
$\delta_{\vec p}=\delta^{(2)}(\vec p_1+\vec p_2-\vec p_{1'}-\vec p_{2'})
$ 
account for the energy and momentum conservation. 
The gain and loss contributions 
are related by the reciprocity symmetry $12\leftrightarrow 1'2'$. Here $V_{12,1'2'}$ is the two-body interaction, 
properly antisymmetrized to account for Fermi statistics. 
Interaction $V_{12,1'2'}$ depends on momentum transfer $k$ on the $k\sim k_F$ scale; 
this $k$ dependence is inessential and will be ignored.  
In what follows we consider a spatially uniform problem setting $[f,H]=0$. 
The sum over momenta $2$, $1'$, $2'$ represents a six-dimensional integral over $\vec p_2$, $\vec p_{1'}$ and $\vec p_{2'}$, which is discussed below. 

For a weak perturbation away from equilibrium,  Eq.\eqref{eq:Golden_Rule} linearized by the standard ansatz $f(\vec p)=f_0(\vec p)-\frac{\p f_0}{\p\epsilon}\eta(\vec p)$,
where $f_0(p)$ denotes the equilibrium Fermi distribution,  
yields a linear integrodifferential equation $f_{0}(1-f_{0})\frac{d \eta_1}{dt}
=I_{\rm ee}[\eta]$ with the operator $I_{\rm ee}$ given by
\be\label{eq:I_ee}
I_{\rm ee}[\eta]=\sum_{21'2'} 
\frac{2\pi}{\hbar}|V|^2 F_{121'2'} 
\delta_\epsilon \delta_{\vec p}
\lp \eta_{1'}+\eta_{2'}-\eta_{1}-\eta_{2}\rp 
\ee
Here $\sum_{21'2'}$ and $|V|^2$ 
denote the six-dimensional integral $\int\frac{d^2p_2d^2p_{1'}d^2p_{2'}}{(2\pi)^6}$ and the interaction matrix element $|V_{12,1'2'}|^2$, whereas 
the quantity $F_{121'2'}$ is 
a product of the equilibrium 
Fermi functions
$f^0_{1}f^0_{2}(1-f^0_{1'})(1-f^0_{2'})$. 

Different excitations are described as eigenfunctions of the collision operator $I_{\rm ee}$, 
with the eigenvalues giving the decay rates equal to inverse lifetimes. Because of the cylindrical symmetry of the problem, the eigenfunctions are products of angular harmonics on the Fermi surface and functions of the radial energy variables $x_i=\beta(\epsilon_i-\mu)$:
\be
\eta(\vec p,t)= \sum_m e^{-\gamma_m t} e^{i m\theta}\chi_m(x)
,
\ee
where $\gamma_m$ and $\chi_m(x)$ are solutions of the spectral problem $-\gamma_m 
f_{0}(1-f_{0}) \chi_m(x)=I_{\rm ee}[\chi_m(x)]$. 

 \begin{figure*}[t]
\includegraphics[width=0.7\textwidth]{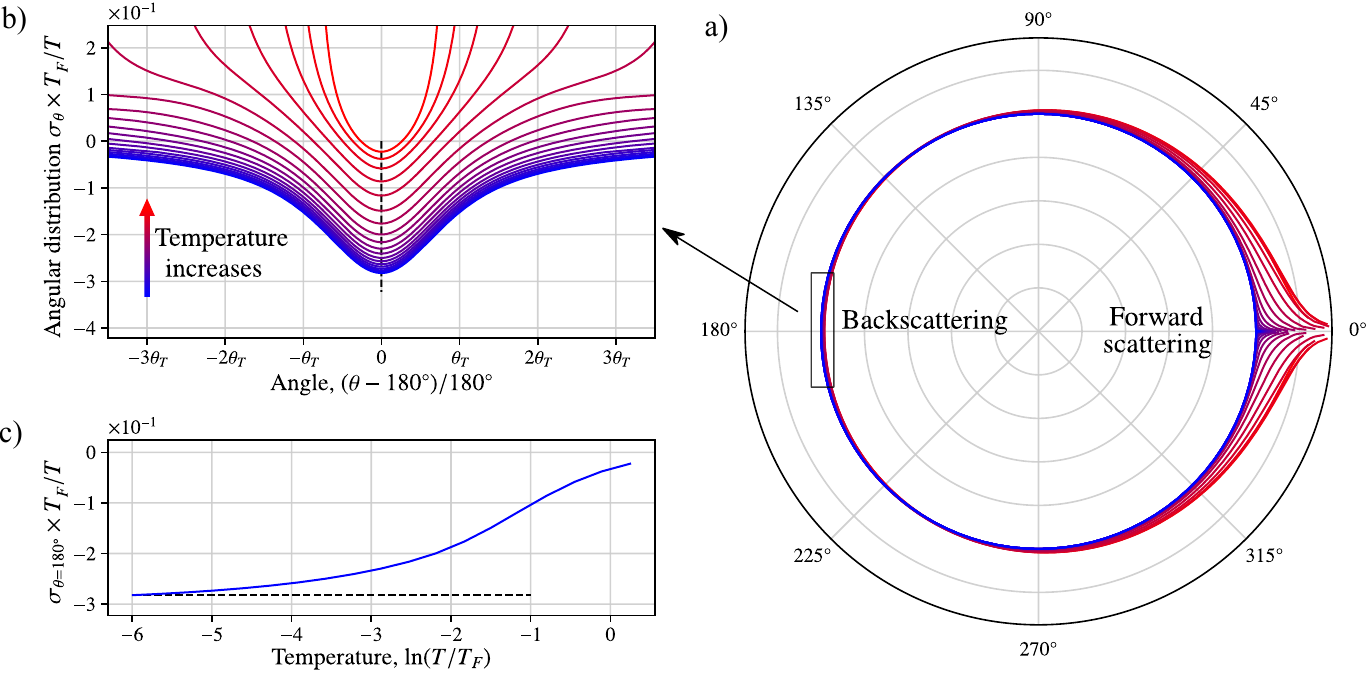}
\caption{
a) Angular distribution $\sigma(\theta)$ for two-body quasiparticle scattering at the Fermi surface, Eq.\eqref{eq:crosssection}, 
at different temperatures. Restricted phase space gives rise to collinear scattering, producing sharp peaks in the forward and backward directions, $\theta=0$ and $\pi$. 
Temperature values used: 
$T/T_F =  10^{-2}\times [0.25, 0.5, 1, 2, 4, 8, 16, 32, 64, 128]$.  
b) The back-scattering peak in $\sigma(\theta)$ near $\theta=\pi$ for the same temperatures as in a). Angle is given in $T$-dependent units $\theta_T = T/T_F$ to illustrate linear $T$ dependence of the peak width. The intensity $\sigma(\theta)$ is multiplied by $T_F/T$ to illustrate linear $T$ dependence of the peak height. 
This translates into $\sim T^2$ scaling for the peak area. c) The dependence of peak height vs. $T$ confirms asymptotic linear scaling at low $T$. 
}
\vspace{-5mm} 
\label{fig:3plots}
\end{figure*}


Before we proceed with diagonalizing the operator $I_{\rm ee}$ we note that one more reason for why the long-lived modes have been missed in the literature 
undoubtedly lies in the difficulty of a direct calculation. 
This problem proves to be quite 
demanding for several reasons. First, the eigenstates of $I_{\rm ee}$ are localized in a peculiar phase space region, 
an annulus at the Fermi surface of width proportional to $T$ owing to the fermion exclusion effects (see Sec. A in Supplemental Information \cite{Supplement}). 
Sampling this ``active part'' of $p$ space requires a mesh which is adjusted with temperature. Second, capturing the kinematic constraints that lead to collinear collision effects, requires ``high-finess'' sampling of the near-collinear momenta 
as compared to the generic momenta in the annulus (see Sec. B in Supplemental Information \cite{Supplement}). Things are made still more complex by the fact that the anglular width of the active collinear region also varies with temperature, decreasing as $T$. 
To tackle this problem, we make use of the cylindrical symmetry of our system and link the decay rates for different modes to the angular distribution for scattering induced by a test particle injected in the system. Computing the angular distribution as described below, we Fourier-transform it in $\theta$ 
to find decay rates for individual modes. This scheme allows us to directly diagonalize the collision operator, Eq.\eqref{eq:I_ee}, finding the results shown in Fig.\ref{fig:fig1} 
(the relevant technical steps are described in Secs. C and D Supplemental Information). 

The angular distribution of particles scattered after  a test particle has been injected in the system at an energy near the Fermi level,
$f_i(\theta)=J_0\delta(\theta-\theta_i)$, is given by
\be\label{eq:crosssection}
f(\theta)=\oint \frac{d\theta'}{2\pi} \sigma(\theta-\theta')f_i(\theta')
=\frac{J_0}{2\pi}
\sigma(\theta-\theta_i)
,
\ee
where $f_i(\theta)$ 
describes the injected beam and the scattering angle $\theta$ parameterizes the Fermi surface. Here $J_0$ is a $T$-independent intensity of the injected beam and, for simplicity, we suppressed the width of the distribution in the radial direction. As discussed above, excitations with different lifetimes are represented as normal modes of the two-body collision operator linearized in the deviation of the distribution from the equilibrium state
$
I_{\rm ee} [f_m(\theta)]=-\gamma_m f_m(\theta)
$,
where $\gamma_m$ are the decay rates (inverse lifetimes) for different excitations. 
Due to the cylindrical symmetry of the problem, the normal modes are  the angular harmonics $f_m(\theta)=e^{im\theta}$ times some functions of the radial momentum variable \cite{Supplement}. Comparing to Eq.\ref{eq:crosssection} we see that the quantities $\gamma_m$ are related to the Fourier coefficients of the angle-resolved cross-section, 
\be\label{eq:sigma(theta)_harmonics}
{\textstyle \sigma(\theta)=\sum_m  e^{im(\theta-\theta_i)}(\gamma_m-\gamma_0)
,}
\ee
where the term $-\gamma_0$ describes particle loss from the injected beam. 
We use the basis functions introduced above to compute $\sigma(\theta)$ and then use the relation in \eqref{eq:sigma(theta)_harmonics} 
to obtain lifetimes of different modes. 

The angular dependence, shown in Fig.\ref{fig:3plots}, features sharp peaks centered at $\theta=0$ and $\pi$, describing forward scattering and backscattering, respectively. The angular widths $\theta_T$ of the peaks scale as $T$ at $T\ll T_F$. Notably, the backscattering peak is of a negative sign, representing backreflected holes. At $T\ll T_F$ the values $\sigma(\theta)$ at generic $\theta$ within the peak scale as $T$. Multiplying this by the peak width 
$\theta_T\sim T/T_F$ yields the net backscattering rate that scales as $T^2/T_F$, as expected from Fermi-liquid theory. This behavior is detailed in Fig.\ref{fig:3plots} insets. 

The decay rates $\gamma_m$ for odd-$m$ modes, obtained from the relation in \eqref{eq:sigma(theta)_harmonics}, show
significant departure from a $T^2$ scaling. The even-$m$ and odd-$m$ rates, 
shown in Fig.\ref{fig:fig1}, are similar at $T\sim T_F$ but have a very different behavior at $T<T_F$. This difference 
originates from the collinear character of scattering, manifest in prominent peaks in $\sigma(\theta)$ in the forward and backward directions. The near-equal areas of these peaks and the negative sign of the backscattering peak suppress the odd-$m$ Fourier harmonics of $\sigma(\theta)$, yielding small decay rates for these harmonics. 
The $T$ dependence for the even-$m$ harmonics agrees well with the $T^2$ law. The odd-$m$ harmonics, to the contrary, have decay rates decreasing at low $T$ much faster than $T^2$. For these harmonics, we observe scaling $\gamma_m\sim T^\alpha$ with $\alpha$ slightly below 4. This represents 
a ``super-Fermi-liquid'' suppression of the decay rates for odd-$m$ harmonics. 

It is interesting to mention that collinear scattering, manifest in the sharp peaks in 
$\sigma(\theta)$ at $\theta=0$ and $\pi$, is directly responsible for the log enhancement of quasiparticle decay rates predicted from the self-energy analysis
\cite{Hodges1971,Chaplik1971,Bloom1975,Giuliani1982,Zheng1996,Menashe1996,
Chubukov2003}. Indeed the angle dependence 
near  $\theta=0$ and $\pi$ is of the form $\sigma(\theta)\sim T^2/|\theta|$ and $T^2/|\theta-\pi|$, with the $1/|\theta|$ singularity rounded on the scale $\delta \theta\sim T/T_F$, as illustrated in Fig.\ref{fig:3plots}. 
Integrating the angle-resolved crosssection over $\theta$ yields a $\log(T_F/T) T^2$ total scattering crosssection. 
This illustrates that 
the abnormally long-lived excitations with the decay rates that scale as $T^4$ rather than $T^2$, described in this work, and the seminal $\log(T_F/T) T^2$ decay rates
\cite{Hodges1971,Chaplik1971,Bloom1975,Giuliani1982,Zheng1996,Menashe1996,
Chubukov2003}, originate from the same phase-space constraints. Restricted phase space renders quasiparticle scattering a highly collinear process even when the microscopic interactions have a weak angular dependence. 

Given these findings, there is a clear interest to find a simple explanation for the unique properties of long-lived excitations. To accomplish this, we have employed a clever method developed 50 years ago in Refs.\cite{Brooker1968,Jensen1968,Sykes1970,Baym1991} to tackle transport in 3D Fermi liquids. This approach involves linearizing the kinetic equation near thermal equilibrium at $T\ll T_F$ to transform it into a time-dependent Schroedinger equation with a reflectionless ${\rm sech^2}$ potential, which can be solved exactly to predict transport coefficients at $T\ll T_F$. We use this framework to explore the modification of this equation in the 2D case and find that, although the decay rates of most excitations follow the $T^2$ scaling, a unique set of non-decaying excitations emerge due to zero modes originating from the supersymmetric quantum mechanics, with one mode per each odd angular momentum.

The six-dimensional integral operator $I_{\rm ee}$ in Eq.\eqref{eq:I_ee} has a complicated structure which in a general case may be difficult to analyze. However, at $T\ll T_F$ the part of phase space in which transitions $12\leftrightarrow 1'2'$ are 
not restricted by fermion exclusion is a thin annulus of radius $p_F$ and a small thickness $\delta p\approx T/v\ll p_F$. 
One can therefore 
factorize the six-dimensional integration over $\vec p_2$, $\vec p_{1'}$ and $\vec p_{2'}$ in $I_{\rm ee}$ into a three-dimensional energy integral and a three-dimensional angular integral, and integrate over angles to obtain a closed-form equation for the radial dependence $\chi(x)$. This is done by noting that the delta functions $\delta_\epsilon \delta_{\vec p}$ together with the conditions $|\vec p_1|\approx |\vec p_2|\approx |\vec p_{1'}|\approx |\vec p_{2'}|\approx p_F$ imply that the 
states $1$, $2$, $1'$ and $2'$ form two anti-collinear pairs 
\be\label{eq:anticollinear_pairs}
\vec p_1+\vec p_2\approx 0,\quad
\vec p_{1'}+\vec p_{2'}\approx 0
\ee
The azimuthal angles therefore obey $\theta_1\approx\theta_2+\pi$, $\theta_{1'}\approx\theta_{2'}+\pi$. In a thin-shell approximation $\delta p\ll p_F$, this gives two delta functions $\delta(\theta_1-\theta_2-\pi)$, $\delta(\theta_{1'}-\theta_{2'}-\pi)$ that cancel two out of three angle integrals in  $I_{\rm ee}$, allowing to rewrite the quantity $\eta_{1'}+\eta_{2'}-\eta_{1}-\eta_{2}$ as
\be\label{eq:eta+(-)^m eta}
e^{im\theta_{1'}}(\chi(x_{1'})+(-)^m\chi(x_{2'}))
-e^{im\theta_{1}}(\chi(x_{1})+(-)^m\chi(x_{2}))
,
\ee
where $\chi$ is a shorthand for $\chi_m$. Here, as above, the variables $x_i$ denote particle energies scaled by temperature, $x=\beta(\epsilon_i-\mu)$. 
Subsequent steps differ for the even and odd $m$, because the contributions of $\chi(x_{1'})$ and $\chi(x_{2'})$ to $I_{\rm ee}$ cancel out for odd $m$ and double for even $m$, since the quantity $F$  in Eq.\eqref{eq:I_ee} is symmetric in $x_{1'}$ and $x_{2'}$. Focusing on the odd $m$ and carrying out integration over the angle between $\vec p_1$ and $\vec p_{1'}$ yields 
\be\label{eq:chi(x)_integral_eqn}
\tilde F \frac{d\chi(x_1)}{dt}=T^2\int dx_2 dx_{1'} dx_{2'} F
g\delta_x [\chi(x_1)-\chi(x_2)]
,
\ee
where $\tilde F=f_{0}(1-f_{0})$ and $\delta_x=\delta(x_1+x_2-x_{1'}-x_{2'})$. 
Here $T^2$ originates from nondimensionalizing the energy variables $x_i$ in the integral and the delta function, the dimensionless factor $g$ is a result of angular integration, the quantity $F$ 
is defined above. Integration over energy variables $x_2$, $x_{1'}$, $x_{2'}$ extends throughout $-\infty<x_i<\infty$, as appropriate for $T\ll T_F$. 

As a first step, we reverse signs of the $1'$ and $2'$ variables: $x_{1'}\to -x_{1'}$, $x_{2'}\to -x_{2'}$. This transforms the integral equation in Eq.\eqref{eq:chi(x)_integral_eqn} to
\begin{align}\label{eq:chi(x)_integral_eqn_no_delta}
& \tilde F \frac{d \chi}{dt} = gT^2 \int dx_2 dx_{1'}dx_{2'} F_{121'2'} \delta_x^+(\chi(x_1) - \chi(x_2)), \nonumber
\\
& F_{121'2'}=f_0(x_1)f_0(x_2)f_0(x_{1'})f_0(x_{2'})
\end{align}
where $\delta_x^+=\delta(x_1+x_2+x_{1'}+x_{2'})$. 
Next we use the identities
\begin{align}
&\int dx_2 dx_{1'}dx_{2'}f_0(x_2)f_0(x_{1'})f_0(x_{2'})\delta_x^+
=\frac12\frac{x_1^2+\pi^2}{1+e^{-x_1}}
,\\
&\int dx_{1'}dx_{2'}f_0(x_{1'})f_0(x_{2'})\delta_x^+
=-\frac{x_1+x_2}{1-e^{-x_1-x_2}}
\end{align}
to carry out integration over $x_2$, $x_{1'}$, $x_{2'}$ in the first term and over $x_{1'}$, $x_{2'}$ in the second term. 
The equation can be further simplified using the substitution
\begin{equation}\label{eq:chi_substitution}
\chi(x) = 2 \cosh \left( \frac{x}{2} \right) \zeta(x)=\lp e^{x/2}+e^{-x/2}\rp \zeta(x),
\end{equation}
which gives an equation
\[
\frac{d\zeta(x_1)}{dt}=-gT^2\lb \frac{x_1^2+\pi^2}2\zeta(x_1)+\int dx_2 \frac{\bar x}{\sinh\bar x}
\zeta(x_2)\rb
,
\]
where $\bar x=(x_1+x_2)/2$. 
Next, we reverse the sign of $x_2$, which brings the integral operator to the form of a convolution, separately for the even and odd functions $\zeta(x_2)$. For an even function $\zeta(-x_2)=\zeta(x_2)$ we have
\[
\int dx_2 \frac{x_1-x_2}{2\sinh\frac{x_1-x_2}2}\zeta(x_2)
.
\]
After Fourier transform $\zeta(x)=\int dk e^{ikx}\psi(k)$ this gives a time-dependent Schroedinger equation with a ${\rm sech^2}$ potential, 
\be\label{eq:TDSE}
\p_t\psi(k)=gT^2\lb \frac12 \psi''(k) -\lp \frac{\pi^2}2- \frac{\pi^2}{\cosh^2\pi k}\rp \psi(k) \rb
.
\ee
Unlike the 3D case, where after a similar transformation the $T^2$ scaling translates into a $T^2$ dependence of the decay rates, here the operator in \eqref{eq:TDSE} has a zero mode, $\psi_0(k)=\frac1{\cosh(\pi k)}$. Being a zero mode, this mode does not relax. 
The associated $\chi_0(x)$  can be found from the identity $\int d\xi\frac{e^{2\pi i \xi y}}{\cosh \pi \xi}=\frac1{\cosh \pi y}$, giving $\chi_0(x)=1$. Returning to the energy variable, this yields the Fermi-surface-displacement mode $\delta f(x)=df_0/dx=f_0(1-f_0)$, identical for all odd $m$. 

In a similar manner, for odd functions $\zeta(-x_2)=-\zeta(x_2)$ upon changing $x_2$ to $-x_2$ a minus sign appears in front of the integral operator:
\[
-\int dx_2 \frac{x_1-x_2}{2\sinh\frac{x_1-x_2}2}\zeta(x_2)
.
\]
Carrying out Fourier transform 
$\zeta(x)=\int dk e^{ikx}\psi(k)$ give a time-dependent Schroedinger equation for 
a ${\rm sech^2}$ potential of an opposite sign
\be \label{eq:diffeq_psi}
\p_t\psi(k)=gT^2\lb \frac12 \psi''(k) -\lp \frac{\pi^2}2+ \frac{\pi^2}{\cosh^2\pi k}\rp \psi(k) \rb
\ee
In this case, physical solutions correspond to the eigenfunctions that are odd in $k$. For a repulsive ${\rm sech^2}$ potential these functions are in the continuum spectrum and asymptotically have the form of plane waves.  As a result, the behavior of the eigenfunctions that are odd in $x$ is quite different from that of the even-$x$ eigenfunctions discussed above. 

For even values of $m$, the analysis follows a similar approach, resulting in analytical expressions for the eigenstates and corresponding eigenvalues. However, the 1D Schrödinger operators associated with even $m$ exhibit no zero modes. 
As a result, the analysis yields a normal $T^2$ scaling of the decay rates. This is so because for even $m$ the terms $\chi(x_1')$ and $\chi(x_2')$ in \eqref{eq:eta+(-)^m eta} are of equal signs and do not cancel out. As a result, the even-$m$ and odd-$m$ harmonics show a very different behavior: 
the odd-$m$ rates vanish in the zero-thickness approximation for the active shell at the Fermi surface, whereas the even-$m$ rates remain finite in this limit, scaling as $T^2$. 

We would like to note that, while the 1D quantum mechanics treatment of
even-$m$ excitations is complete, the corresponding problem for odd-$m$ excitations remains open and requires further investigation. 
Infinite lifetimes found for odd-$m$ modes and interpreted in terms of zero modes, merely indicate that the decay rates for these modes vanish at order $T^2$. 
However, the supersymmetry that protects zero eigenvalues is a property that only appears in the limit of zero thickness of the thermally broadened Fermi surface. Therefore, it is unlikely that this property holds outside of this limit, and we expect the lifetimes of odd-$m$ modes to be finite. Our numerical results indicate 
that the decay rates for these modes scale as $T^\alpha$, with $\alpha>2$. However, determining the values of $\alpha$ analytically may require a framework that extends beyond the approximations considered in our 1D quantum mechanics approach. 

Further research is needed to understand the behavior of odd-$m$ excitations, and we hope that our work will inspire future forays into this intriguing problem. 
The relation with the 1D supersymmetric quantum mechanics can be employed, in principle, to study a variety of other problems of interest, e.g. the thermal transport effects such as thermal conduction, the Joule-Thomson effect and convective thermal drag. 
A comprehensive understanding of these 
transport effects would require deriving transport equations for these quantities supplied with suitable boundary conditions and connecting them to observables. This is an interesting topic for future work.

In summary, the kinematic restrictions of the phase space for quasiparticle scattering at the Fermi surface lead to highly collinear dynamics, even if the microscopic interactions have weak angular dependence. This gives rise to several notable effects, such as the emergence of abnormally long-lived excitations and strong backscattering features in the angular distribution for  two-body collisions. The resulting unusual kinetics is especially relevant for 2D systems that are currently being investigated for electron hydrodynamics and related collective phenomena. 

Long-lived degrees of freedom can amplify the response to weak perturbations, giving rise to long-lasting collective memory effects and new hydrodynamic modes. This is illustrated by a family of viscous modes with non-Newtonian viscosity and transport phenomena due to these modes described in \cite{Kryhin2023} This area of transport theory is rapidly evolving, and a robust understanding of the fundamental physics behind collinear collisions is crucial to grasp the electron behavior in various transport phenomena.

We thank Dmitry Maslov for inspiring discussions and Rokas Veitas for assistance at the initial stages of this project. This work was supported by the Science and Technology Center for Integrated Quantum Materials, NSF Grant No.\,DMR1231319; Army Research Office Grant W911NF-18-1-0116; US-Israel Binational Science Foundation Grant No.\,2018033; and Bose Foundation Research fellowship.




\phantom{}
\newpage 
\phantom{}
 \pagebreak

\begin{widetext}
	\begin{center}{\bf Supporting Material for ``Collinear scattering and long-lived excitations in two-dimensional electron liquids"}
	\author{Serhii Kryhin and Leonid Levitov}
	\end{center}
	\maketitle
	
	This supplemental information describes the essential steps of the numerical analysis that has led to the results presented in Fig. 1 of the main text.  
Sec. A 
outlines the analytical steps used to bring the linearized collision operator to the form suitable for 
performing numerical integration. Sec. B defines the basis used for numerical computation and the choice of a mesh in $k$ space. Sec. C describes matrix representation of the linearized collision operator 
and its relation to the angular distribution and other quanitites of interest considered in the main text. Sec. D assesses limitations of the present approach and discusses several numerical instabilities we had to overcome in our analysis.

\subsection{Direct diagonalization of the linearized collision operator: an overview}

Here we 
describe how the collision integral 
is simplified and brought to the form amenable to 
numerical integration.
We proceed in two steps, first using the kinematic constraints to reduce the six-dimensional integral in 
Eq. \eqref{eq:I_ee} 
of the main text to a three-dimensional integral, and then using a suitable basis of functions to reduce the three-dimensional integrals to one-dimensional integrals. After that, the operator $I_{\rm ee}$ can be projected on a subspace that represents adequately the states on the active shell and diagonalized numerically. 


Integration over $\vec p_2$ can be eliminated by a momentum-conservation delta-function, giving
\be\label{eq:appendix_Iee}
    I[\eta]
    =  -
    \frac{2\pi}{\hbar} |V|^2
    \int \frac{d^2 p_{1'} \, d^2 p_{2'}}{(2\pi)^4} F_{121'2'} 
\delta_\epsilon 
\cdot {\sum}'_\alpha \eta_\alpha
,
\ee
where $\vec p_2$ is now a function of the other momenta,
$\vec p_2 = \vec p_{1'} + \vec p_{2'} - \vec p_1$. As above, $\delta_\epsilon$ denotes $\delta(\epsilon_1+\epsilon_2-\epsilon_{1'}-\epsilon_{2'})$ and ${\sum}'_\alpha \eta_\alpha$ stands for $\eta_{1'}+\eta_{2'}-\eta_1-\eta_2$. 
Next, we eliminate the radial integration over $|\vec p_{2'}|$ by canceling it with the energy delta-function. 
The expression for the collision integral then takes the form 
\begin{equation}
I[\eta]
= -A 
\int \frac{d^2 p_2 \,  d\theta_{\vec{n}}  }{(2\pi)^4} 
F_{121'2'}
{\sum}'_\alpha \eta_\alpha,
\quad
A = \pi m |V|^2 / \hbar^3
,
\label{eq:int_for_calc}
\end{equation} 
where we introduced
 an angle $\theta_{\vec{n}}$ defined through the kinematic relation between the incoming and outgoing momenta as follows. 
%
Due to momentum and energy conservation, in
Eq. \eqref{eq:int_for_calc} the momenta $\vec p_{1'}$ and $\vec p_{2'}$
satisfy the constraints
\be\label{eq:const_pipjpr}
\vec p_{1'} =\vec p_{+}+|\vec p_{-}|\vec n,\quad
\vec p_{2'} =\vec p_{+}-|\vec p_{-}|\vec n,
\ee
where $\vec p_{\pm}=\frac{\vec p_1 \pm \vec p_2}2$ and we introduced a unit vector $\vec{n} = (\cos \theta_{\vec{n}}, \sin \theta_{\vec{n}})$ that parameterizes the outgoing momenta in the collision process, wherein the incoming momenta $\vec p_1$ and $\vec p_2$ are taken to be fixed, see schematic in Fig.\ref{fig:n_vec}.

The temperature-independent constant $A$ 
defined in Eq.\ref{eq:int_for_calc}, which describes the scattering potential strength, can be used to non-dimensionalize the decay rates for different excitations, $\gamma_m$. 
Namely, $\gamma_m$ are related to the eigenvalues  $\lambda_m$ of the dimensionless linear integral operator introduced in the main text and analyzed below as
\be\label{eq:lambda_m_gamma_m}
\gamma_m=Ap_F^2\lambda_m
\ee 
(as a reminder, $p_F$ denotes the Fermi wavevector rather than the Fermi momentum, and thus has units 1/length). This relation can be used to relate the eigenvalues $\lambda_m$ shown in Fig. 1 of the main text with physical decay rates. 

\begin{figure}[t]
\centering
\includegraphics[width=0.4\columnwidth]{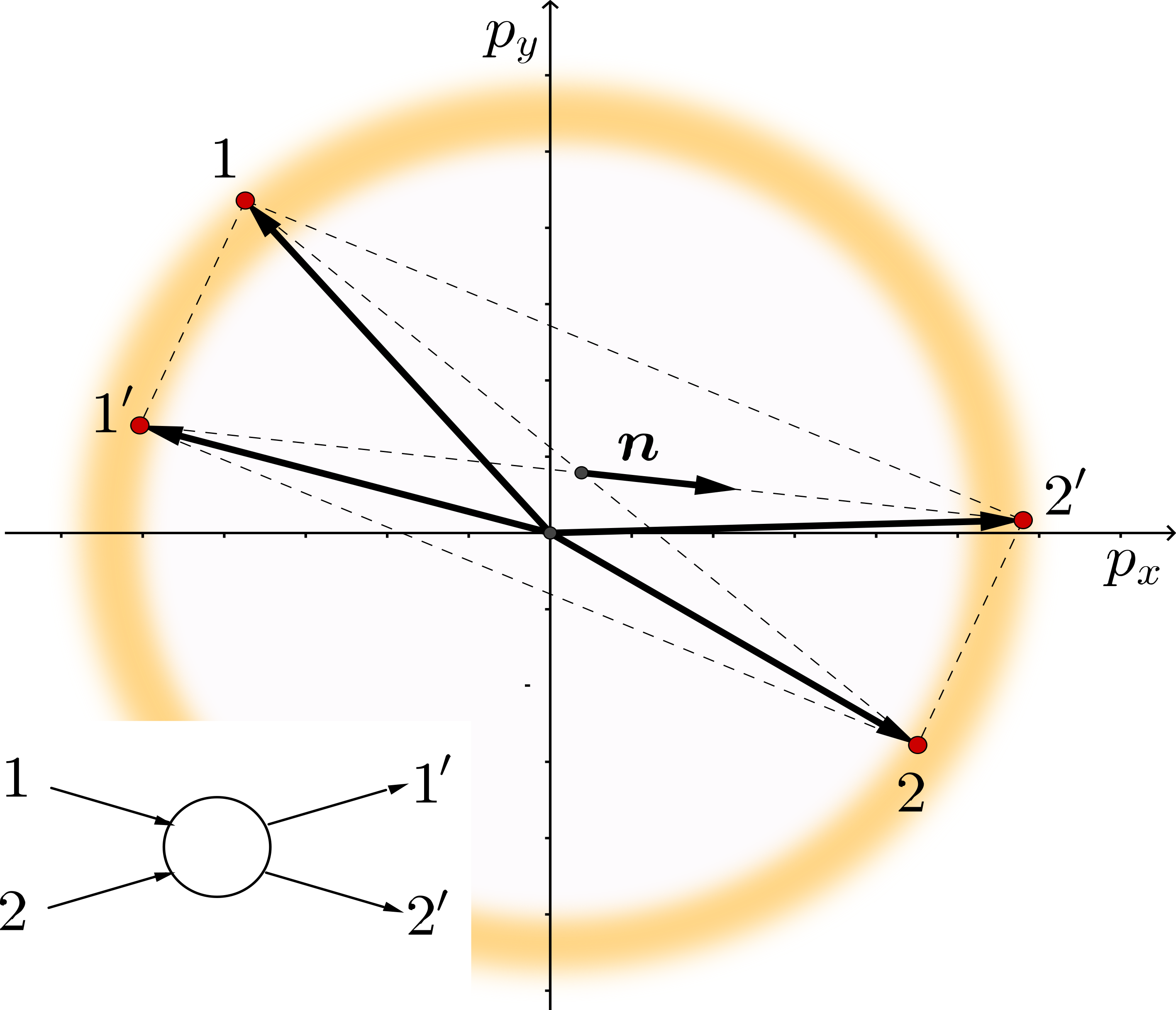}
\caption{Ingoing and outgoing momenta that contribute to excitation dynamics for  a typical scattering processes $12\to 1'2'$ shown in the inset. The blurred annulus of radius $p=p_F$ and width $\delta p\sim T/v$ is the region near the Fermi surface where collisions are allowed by fermion exclusion. Kinematic constraints 
select processes in which momenta form nearly anticollinear pairs 1-2 and $1'$-$2'$, see 
Eq. 7 of the main text. Shown is the vector $\vec{n}$, \eqref{eq:const_pipjpr}, used to parameterize momentum states, 
$\vec{n} =\frac{\vec p_{2'}-\vec p_{1'}}{|\vec p_{2'}-\vec p_{1'}|}= (\cos \theta_{\vec{n}}, \sin \theta_{\vec{n}})$. 
}
\label{fig:n_vec}
\end{figure}


Next, we choose a basis of functions to represent the states $\eta(\vec p_1)$ and define a matrix representation for the linear operator $I[\eta_{\vec p_1}]$. 
Different choices of basis functions have different computational 
limitations. 
Here we employ, as a basis, the $\delta$-functions
\be\label{eq:eta_delta_function}
\eta_k(\vec p) = \delta^{(2)}(\vec p - \vec k)
\ee 
labeled by different $\vec k$ [for a discussion of normalization, which depends on the choice of the mesh, see  Sec. B and C below]
This basis combines computational efficiency with analytic simplicity. 
Indeed, the two-dimensional delta functions, \eqref{eq:eta_delta_function}, when substituted in the collision operator, cancel two out of three integrations in $\int dp_2 d\theta_2 d\theta_{\vec{n}}...$ yielding an expression that involves just one integral. 

An added benefit of working in the delta-function basis is that it allows to analytically simplify the expression for the collision operator, \eqref{eq:int_for_calc}. The collision operator can be written as a sum of four contributions, one for each $\eta_\alpha$. This yields an expression $I[\eta]$ = $I_1[\eta]$ + $I_2[\eta]$ - $I_3[\eta]$ - $I_4[\eta]$ with the individual terms given below: 
\begin{align}
	I_1[\eta_k] =& -A 
	\eta_k(\vec p_1) \int \frac{d^2 p_2 \, d\theta_{\vec{n}}}{(2\pi)^4} F_{121'2'},
	\label{eq:int_1} 
	\\
	I_2[\eta_k] = & - A 
	\int \frac{d\theta_{\vec{n}}  }{(2\pi)^4} F_{121'2'} ,
	\label{eq:int_2}
\end{align}
where $I_1$ and $I_2$ represent the contributions of $\eta_1$ and $\eta_2$ 
in \eqref{eq:int_for_calc} and $F$, as above, denotes $F=f^0_1 f^0_2 (1-f^0_{1'}) (1 - f^0_{2'})$. 
In \eqref{eq:int_2} we eliminated two integrals by integrating
over a delta-function. Integrals $I_3$ and $I_4$, which correspond to $\eta_{1'}$ and $\eta_{2'}$ respectively, can be written in a similar way: 
\begin{align}
I_3[\eta_k] & = -A 
\int \frac{d^2 p_2 \, d\theta_{\vec{n}}}{(2\pi)^4} F_{121'2'} \,  
\eta_k(\vec p_{1'}),
\label{eq:I3}
\\
I_4[\eta_k] & = -A 
\int \frac{d^2 p_2 \,  d\theta_{\vec{n}}  }{(2\pi)^4} F_{121'2'} \, 
\eta_k(\vec p_{2'}).
\label{eq:I4}
\end{align}
In this case, eliminating integration over $\vec p_2$ by canceling it with the delta functions $\eta_k(\vec p_{1'})$, $\eta_k(\vec p_{2'})$ 
is a little more cumbersome. 
In the term $I_4$ the $\delta$-function constraint is 
\begin{equation}
	\vec p_{2'}(\vec p_1, \vec p_2, \theta_{\vec{n}}) = \vec k,
	\label{eq:delta_const}
\end{equation}
where the expression for $\vec p_{2'}$ is given in \eqref{eq:const_pipjpr}. This equation should be solved for $\vec p_2(\vec p_1, \vec k, \theta_\vec{n})$, which is a zero of the $\delta$-function's argument. 
To perform integration over $\vec p_2$ in \eqref{eq:I4} we use the value $\vec p_{1'}(\vec p_1, \vec k, \theta_\vec{n})$ and evaluate the Jacobian at the zero of the delta function. 
Conveniently, Eq. \eqref{eq:delta_const} can be solved in a closed form, after which the first relation in \eqref{eq:const_pipjpr} yields
\begin{align}
    \vec p_2(\vec p_1, \vec k, \theta_{\vec n}) = 2 \vec k - \vec p_1 - \frac{(\vec k - \vec p_1)^2 \; {\vec n}}{(\vec k - \vec p_1) \cdot {\vec n}}
    \label{eq:sub_pj}
    \\
    \vec p_{1'}(\vec p_1, \vec k, \theta_{\vec n}) = \vec k - \frac{(\vec k - \vec p_1)^2 \; {\vec n}}{(\vec k - \vec p_1) \cdot {\vec n}}
    \label{eq:sub_piprime}.
\end{align}
The Jacobian of a $\delta$-finction in $I_4$ is
\begin{equation}\label{jacobian}
    J = \frac{(\p p_{2'}^x,\p p_{2'}^y)}{(\p p_2^x,\p p_2^y)}= \frac{1}{2} \frac{((\vec k - \vec p_1) \cdot \vec{n})^2}{(\vec k - \vec p_1)^2}
\end{equation}
(found by linearizing the second relation in \eqref{eq:const_pipjpr}).
After handling the contribution $I_3$ in a similar manner, the sum of $I_3$ and $I_4$ can be simplified to read
\be
   (I_3+I_4)[\eta_k] =  -A
    \int \frac{d\theta_{\vec{n}}}{(2\pi)^4} \, J^{-1} F_{121'2'},
    \label{eq:int_34}
\ee
where $\vec p_2$ and $\vec p_{1'}$ are given by Eqs.\eqref{eq:sub_pj} and \eqref{eq:sub_piprime}, and the Jacobian $J$ is given by \eqref{jacobian}.

Importantly, evaluating $I_2$, $I_3$ and $I_4$ on a delta function state, Eq.\ref{eq:eta_delta_function}, yields smooth functions of $\vec p$ given by simple 1D  integrals.
The $I_1$ contribution, to the contrary, yields a delta function identical to the one in \eqref{eq:eta_delta_function}, with a prefactor that is given by a 3D integral. This contribution describes particle loss from the initial state $\vec p_1$, the contributions $I_2$, $I_3$ and $I_4$ describe gain.

One peculiar aspect of working with delta functions is a Jacobian that has a non-analytic structure, \eqref{jacobian}.
We note that, while the Jacobian $J$ is zero when condition $(\vec k - \vec{p_1})\cdot \vec{n} = 0$ is satisfied, 
this does not mean that the whole expression inside the integral is divergent. The behavior of the integral around the divergent points of a phase space can be understood by introducing new variables $\Delta p$ and $\phi$, such that 
$\Delta p = |\vec k - \vec{p}_1|$ and 
$\cos{\phi} = (\vec k - \vec{p}_1)\cdot \vec{n} / \Delta p$.
In these variables the Jacobian $J$ can be written as
\begin{equation}
	J = \cos^2 \phi,
\end{equation}
an expression that remains finite and non-zero so long as $\cos \phi \ne 0$.
Therefore, a divergence  in the integral in \eqref{eq:int_34} might occur only when the quantity $\cos \phi$ vanishes.
On the other hand, expressions for $\vec p_2$ and $\vec p_{1'}$ in Eqs.\eqref{eq:sub_pj} and \eqref{eq:sub_piprime} have a term $(\vec p_1 - \vec k) \cdot \vec{n}$ in their denominators, which is proportional to $\cos \phi$. Therefore at $\cos \phi = 0$ the absolute values of $\vec p_2$ and $\vec p_{1'}$ diverge so that $|\vec p_2| \rightarrow +\infty$ and $|\vec p_{1'}| \rightarrow +\infty$. This divergence leads to an exponential decrease of the $f(\vec p_2)$ term, which cancels the divergence of $J^{-1}$. Therefore, the expression inside the integral has only an isolated discontinuity point rather than a pole and therefore the integral has a finite value. Even though the integral is well-behaved, 
the singularity near $\vec p = \vec k$ point makes the numerical computation problematic. We study the impact of the numerical error in forward scattering below, 
finding that it does not affect the qualitative behavior.

The representation of the collision operator introduced above can be used to 
project it on a subspace spanned by a set of basis functions chosen to provide a sufficiently good sampling of the active region in momentum space (the blurred annulus pictured in Fig.\ref{fig:n_vec}). This yields a finite-size matrix that can be diagonalized to find the excitation eigenmodes and their eigenvalues, giving the decay rates. We have found that, although this direct approach works, it is more convenient to use a somewhat
different approach to the problem, which employs the angular distribution of 
quasiparticle scattering in the active region near the Fermi surface. 

\subsection{The delta-function basis and an optimized integration mesh}




In this section discuss the basis of functions in the momentum space used to represent particle momentum distributions perturbed by collisions. We describe the reasoning behind our basis choice and the integration mesh used in this study. In this work we opted for a basis comprised of suitably normalized delta functions. This choice is quite different from the more conventional approaches relying on systems of orthogonal polynomials (e.g. see \cite{Kanki2011}) The delta functions, being singular functions, may not appear to be a natural choice of a basis. However, in a problem like ours, the delta functions have distinct advantages, since, after being plugged in 
Eq. \eqref{eq:appendix_Iee}
they considerably reduce the number of required integrations. 

As we observed in Eq. \eqref{eq:int_for_calc}, the nature of the $\delta$-functions allows to eliminate two out of three integrations in three out of four terms in plugged in. 
In addition, projection operation is effectively reduced to computing a value of the function of interest in the corresponding point, unlike in continuous functions bases, where we need to compute one more integral to perform the projection on the basis function itself. Additional complication comes from nature of the function inside the integral. At low temperatures it resembles several peaks in its variable space. This makes the Monte Carlo approach to the integration very hard to apply and pushes us to a mesh-based definite integration methods. Let us assume we perform the integration with $M$ points in a mesh in momentum absolute value and with $N$ mesh in angular variables. The computation time of the matrix element in continuous basis takes $O(M^2 N^3)$ time. A non-diagonal element of the matrix, which is formed by only $I_2$, $I_3$, and $I_4$, in some $\delta$-function basis takes just $O(N)$ time. A diagonal element of a $\delta$-function basis, where $I_1$ also has an impact, takes $O(M N^2)$. To solve a linear problem in a $\delta$-function basis, we do not need to compute the matrix elements for each mesh point. As we show below, rotational symmetry of the initial expression allows us to  perform the integration on $M^2 N$ mesh points instead of $M^2 N^2$ for any non-diagonal matrix element and $M$ points instead of $M N$ points for diagonal elements.
As takeaway, rotational symmetry of the initial expression allows to circumvent one of the integrations over $\theta$.
Therefore, the total computation complexity with the $\delta$-function mesh is $O(M^2 N^2)$ instead of $O(M^2 N^3)$ for continuous bases.

With a very specific choice of a smooth basis for this particular problem it is possible to construct a numerical solution of a the same computational complexity. In particular, one needs to be rigorous in choosing the basis in the way that the choice would respect both rotational symmetry and properties of the integral. One of the possible ways to construct such basis is to use the form
\begin{equation}
f_{nm}(\vec p) = P_n(p) e^{i m \theta} \tilde F(p),
\end{equation}
where $p$ and $\theta$ are polar coordinates of $\vec p$, $\tilde F(p) = f (1 - f)$, and $P_n(p)$ is a polynomial of power $n$ which is chosen to make $f_{nm}(\vec p)$ to be orthogonal to $f_{n'm}(\vec p)$ when $n \neq n'$ with respect to the integral inner product. This basis is one of the optimal basises, since it is both rotationally invariant and spans the region around the Fermi surface, but we stick to less complex basis of $\delta$-functions.

Since computation of the collision integral on the delta-functions
is much faster, we will use a sampling of delta-functions as
a subspace basis for computation. 
We define a set of basis vectors as  Kronecker $\delta$-functions
\begin{equation} \label{eq:Matrix_I}
| \vec{p}_i \rangle = \tilde \delta^{(2)}(\vec{p} - \vec{p}_i) \sqrt{\Delta V_i},
\end{equation}
where $\Delta V_i = p_i \, \Delta p_i \, \Delta \theta_i$; $\Delta p_i$ and $\Delta \theta_i$ are sizes of the part of momentum space that corresponds to $i$'th point in polar coordinates. 
By $\tilde \delta^{(2)}(\vec p - \vec p_i)$ in \eqref{eq:Matrix_I} we mean a a function defined on the mesh and that is equal to $0$ when $\vec p \neq \vec p_i$, and is equal to $1/\Delta V_i$ when $\vec p = \vec p_i$. In the limit of dense mesh function $\tilde \delta^{(2)}(\vec p - \vec p_i)$ behaves like a Dirac $\delta$ function. 
The square root of phase space element is added to preserve the normalization of the basis to be $\langle \vec{p}_i | \vec{p}_j \rangle = \delta_{ij}$ with respect to the inner product in the form of an integral over $\vec p$.

We use this basis to represent the operator $I$ as a matrix: 
\begin{equation}
\langle \vec p_i | \; I \; |\vec p_j \rangle \equiv I_{ij} = 
I[\delta^{(2)}(\vec p_i - \vec p_j)] \sqrt{\Delta V_i \, \Delta V_j}.
\end{equation}
By this construction, the expression yields a symmetric matrix. As such it is suitable for computing the angular distribution for two-body scattering, for which the matrix should be applied to a state that represents the incoming state. It should be noted, however, that the eigenvectors and eigenvalues describing different excitations and their lifetimes are not those of the matrix $I$. Rather, they should be obtained from a generalized eigenvalue problem 
$\gamma \tilde F(p) |\psi\rangle=I|\psi\rangle$ with $\tilde F(p)=f_0(1-f_0)$.

On the side, the lowest eigenvalues for each $m$ can be determined more easily from the angular distribution, as discussed in the main text. This approach was used to obtain the eigenvalues shown in Fig. 
\ref{fig:3plots_no_forw} and Fig. \ref{fig:3plots} of the main text. We verified that the direct solution of the generalized eigenvalue problem gives the same eigenvalues, albeit with a lower accuracy. 

Next, we discuss another crucial aspect of our analysis --- sampling of the relevant part of the momentum space. This achieved by constructing a mesh of points on which the delta-function states given in \ref{eq:Matrix_I} are centered. The mesh must have a higher density near the Fermi surface and for near-collinear momenta, and also respect the cylindrical symmetry of the problem. There are several ways through which these requirements can be satisfied. Below we described the approach that proved particularly useful.

To preserve 
the rotational invariance 
of the collision operator, we take the the mesh points on a set of concentric circles centered at $\vec p = 0$, as illustrated in Fig.\ref{fig:lattice}. 
The radial momentum  components  form an equally spaced set of $M$ points in an interval $p_{\min}(T)<p<p_{\max}(T)$ centered at $p=p_F$. To optimize coverage of the phase space within thermally broadened Fermi surface we used temperature dependence of $p_{\min}(T)$ and $p_{\max}(T)$ was optimized 
defined by $\tilde F(p_{\max}, T) = \tilde F(p_{\min}, T) = \alpha$, with $p_{\min} < p_{\max}$ and 
$\alpha$ a small parameter of choice. In this study we used several values of $\alpha$ and $M$ and came to conclusion that the best choice that allows to achieve reasonable precision is $\alpha = 10^{-3}$ and $M = 40$.
In cases when there was no lower-limit solution for $p_{\min} > 0$, the value $p_{\min}$ was set to 0. 
The choice of boundaries on the absolute values of momentum in the mesh
allows us to focus on the physically interesting region of the phase space near a Fermi surface  where $\tilde F(\vec p, T)$ is not exponentially small. For the temperatures $T \sim \epsilon_F$, the sampled region was a disc of the radius 
$\sim \epsilon_F$. For the temperatures $T \ll \epsilon_F$, the sampled region was
an annulus of radius $\epsilon_F$ and thickness of $\sim T$.

We choose a specific mesh point distribution to resemble the properties of the integral as a function of the angle between momenta $\vec p_i$ and $\vec p_j$. To perform the collision operator analysis as a function of the angle, we need to be able to integrate over an absolute value of momentum (i.e. sum over points with the same angular coordinate and different radial coordinates). Because of this, we choose the same angular distribution of points for each circle of constant momentum absolute value. Assuming the $\delta$-function source, we find that at low temperatures most of the scattering is either near-forward scattering or near-back scattering, and the width of the forward and backward peaks scales $\sim T$ at small temperatures. 
To describe this highly anisotropic scattering it is beneficial to define mesh that has a higher density for the angles in the near-forward and near-backward directions. 
To construct a mesh with such properties we choose a uniform mesh in angle to account for the general properties of the angular distribution. To that end, we use a combination 
of a uniform mesh for the angles away from the collinear and anticollinear directions $\theta=0$ and $\pi$ and a denser mesh concentrated in the regions near $\theta \approx 0$ and $\theta \approx \pi$. The width of these two regions is taken to be a function of temperature proportional to $T$, which accounts for the forward and backward scattering distribution becoming sharper as $T$ decreases, as illustrated in Fig. \ref{fig:3plots}. 
of the main text.
The dense forward/backward mesh is taken to be uniform, comprised of $N$ points. The not-so-dense mesh for non-collinear angles is also taken to be uniform, comprised of $N'$ points. This is illustrated in Fig.\ref{fig:lattice}, where the dense and less dense meshes are shown in different colors. In our simulation we used $N=N'=200$. 


For a 2D mesh in momentum space we use a direct product of the radial and angular meshes defined as described above. We denote the mesh points as $| \vec p_{mn} \rangle \equiv | p_m, \theta_n \rangle$, where 
$\vec p_{mn} = (p_m \cos \theta_n, p_m \sin \theta_n)$ and $1<m<M$ and $1<n<N+N'$. 


\subsection{Matrix representation of the linearized collision operator}

In this section we describe in details the method to obtain the angular distribution from the operator projected on the functional basis and show the correspondence of these operations to the operations with original collision integral in the function space.

We choose the source in the form of delta function in an angular space to describe the electron injection along $\theta = 0$.
The radial distribution for injected electrons is chosen to be proportional to $ -\partial f_0 / \partial \epsilon$. This corresponds
to $\eta_0(p_i, \theta_j) \equiv \eta_0(\theta_j) = \delta(\theta - \theta_j)$ when $\theta_j = 0$. To focus on the angular part of the operator,
we contract the matrix with a column corresponding to $\eta_0$; this is equivalent to integrating original expression over the radial coordinates of momenta:
\begin{equation} \label{eq:I_angle_matrix}
\la \theta_i | I | \theta_j \ra \equiv 
\sum_{p, p', \theta, \theta'} 
\la \eta_0(\theta_i) | p, \theta \ra
\la p, \theta | I | p', \theta' \ra
\times
\la p', \theta' | \eta_0(\theta_j) \ra
\end{equation}
Here the summation over $p$ is a summation over all values of $p_m$ where $1 \leq m \leq M$, and summation over $\theta$ is a summation over all values of $\theta_n$ where $1 \leq n \leq N + N'$. The summation over $p$ and $\theta$ in \eqref{eq:I_angle_matrix} corresponds to the integration over momentum space in the following way:
\begin{equation}
\sum_{p, \theta} \la \eta_0(\theta_i) | p, \theta \ra
\la p, \theta | I | p_{m}, \theta_{n} \ra
\quad \leftrightarrow \quad \int_0^{+\infty} dp \; p \int_0^{2 \pi} d\theta \, \eta_0(\theta_i) \, I[\delta^{(2)}(\vec p - \vec p_{m n})],
\end{equation}
Analogous relations can be established for summations over $p'$ and $\theta'$. 

Note that $\la \theta_i | I | \theta_j \ra$ depends only on the angles and does not depend on the absolute values of two momenta anymore. This happened because we assumed that the source of the injected electrons has $-\p f / \p \epsilon$ profile, and we also projected it onto a $- \p f / \p \epsilon$ state. We use this model because we are mainly interested in the decay of the "near-ground-state" modes of 
Eq.\eqref{eq:TDSE} in the main text, which have energy dependence of the form 
$- \p f / \p \epsilon$. The angular distribution of scattered particles $\sigma(\theta_i)$, is obtained by setting $\theta_j = 0$:
\begin{equation}
\sigma(\theta_i) = \la \theta_i | I | 0 \ra,
\end{equation}
The distribution $\sigma(\theta)$ is shown in Fig. \ref{fig:3plots}.
 of the main text.

Initial operator $I[\delta(\vec p - \vec k)](\vec p_1)$ possesses rotational
symmetry in a sense that the integral is only a function of $k = |\vec k|$,
$p_1 = |\vec p_1|$, and an angle in-between $\vec p_1$ and $\vec k$.
Consequently, the matrix  elements $\la \theta_i | I | \theta_j \ra$ depend
only on the $(\theta_i - \theta_j)$ combination: 
$\la \theta_i | I | \theta_j \ra = G(\theta_i - \theta_j)$. 
The eigenvalues of such a matrix are readily obtained by applying a
discrete Fourier transform to $G(\theta)$. Therefore, the eigenvalues of the operator $\la \theta_i | I | \theta_j \ra$
can be obtained in the matrix notation by  transforming $\sigma(\theta)$ as 
\begin{equation} \label{eq:FT_lam}
\lambda_m = \sum_i e^{-i m \theta_i}\sigma(\theta)
\Delta \theta_i.
\end{equation}
The quantity $\sigma(\theta_i)$ has the meaning of the transition rate per unit angle, with the dimensionality of ${\rm sec^{-1} rad}^{-1}$. The dependence $\sigma(\theta)$ is constrained by particle conservation 
\begin{equation}
\sum_i \Delta\theta_i \sigma(\theta_i)=0
\end{equation}
and momentum conservation
\begin{equation}
\sum_i \Delta\theta_i \cos\theta_i \sigma(\theta_i)=\sum_i \Delta\theta_i \sin\theta_i \sigma(\theta_i)=0.
\end{equation}
in two-body collisions. These constraints yield the identities 
\begin{equation}
\lambda_0 = \lambda_1 = 0.
\end{equation}
The accuracy with which these relations hold provides a useful check for the  precision of our numerical method. As an illustration, the eigenvalue $\lambda_1$ is shown in Fig. \ref{fig:fig1} 
 of the main text (dashed curve). 
In general $\sigma(\theta)$ is a sign-changing function, with $\sigma(\theta)<0$ corresponding to the emission of holes in the backward direction. This behavior is illustrated in Fig. \ref{fig:3plots} 
 of the main text.

On the side, because of the rotational symmetry 
we do not need to compute all the entries of the matrix 
$\la p_m, \theta_n | I | p_{m'}, \theta_{n'} \ra$.
For our purpose it is sufficient to evaluate the vector $\la p_m, \theta_n | I | p_{m'}, 0 \ra$, a quantity that for a rotationally invariant problem contains all the
information about the operator and its eigenvalues. Therefore, the computation complexity of the problem is reduced from being quadratic in the number of angular points $N$ to that linear in $N$, while remaining quadratic in the number of radial points $M$.

\begin{figure}[h]
	\centering
	\includegraphics[width=0.45\textwidth]{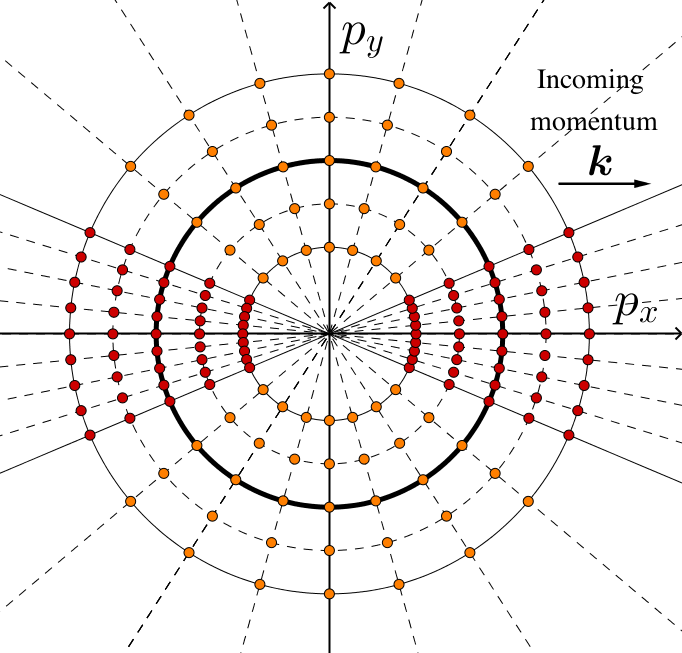}
	\caption{
		The mesh in momentum space used in the calculation. The mesh density is nonuniform to achieve better coverage of angles for the collinear and anti-collinear directions relative to the incoming momentum $\vec k$. This choice guarantees that the mesh respects rotational symmetry of the problem. In the actual calculation we used $M=40$ radial points, $N=200$ azimuthal points for a less dense mesh (orange points), and $N' = 200$ points for a more dense mesh in the collinear and anti-collinear direction of $\vec k$ (red points). Accordingly, the total numbers of angles in the collinear and anti-collinear groups was $N'/2$; the total number angles in the non-collinear upper and lower groups was $N/2$. The radial mesh was chosen to span an annulus covering the Fermi surface (marked by a bold circle). To account for the strong collinear and anti-collinear contributions in the two-body scattering, the $x$, $y$ coordinates are rotated so that the $x$ axis is aligned with the incoming momentum $\vec k$. }
	\label{fig:lattice}
\end{figure}

\begin{figure}[t]
	\centering
	\includegraphics[width=0.7\textwidth]{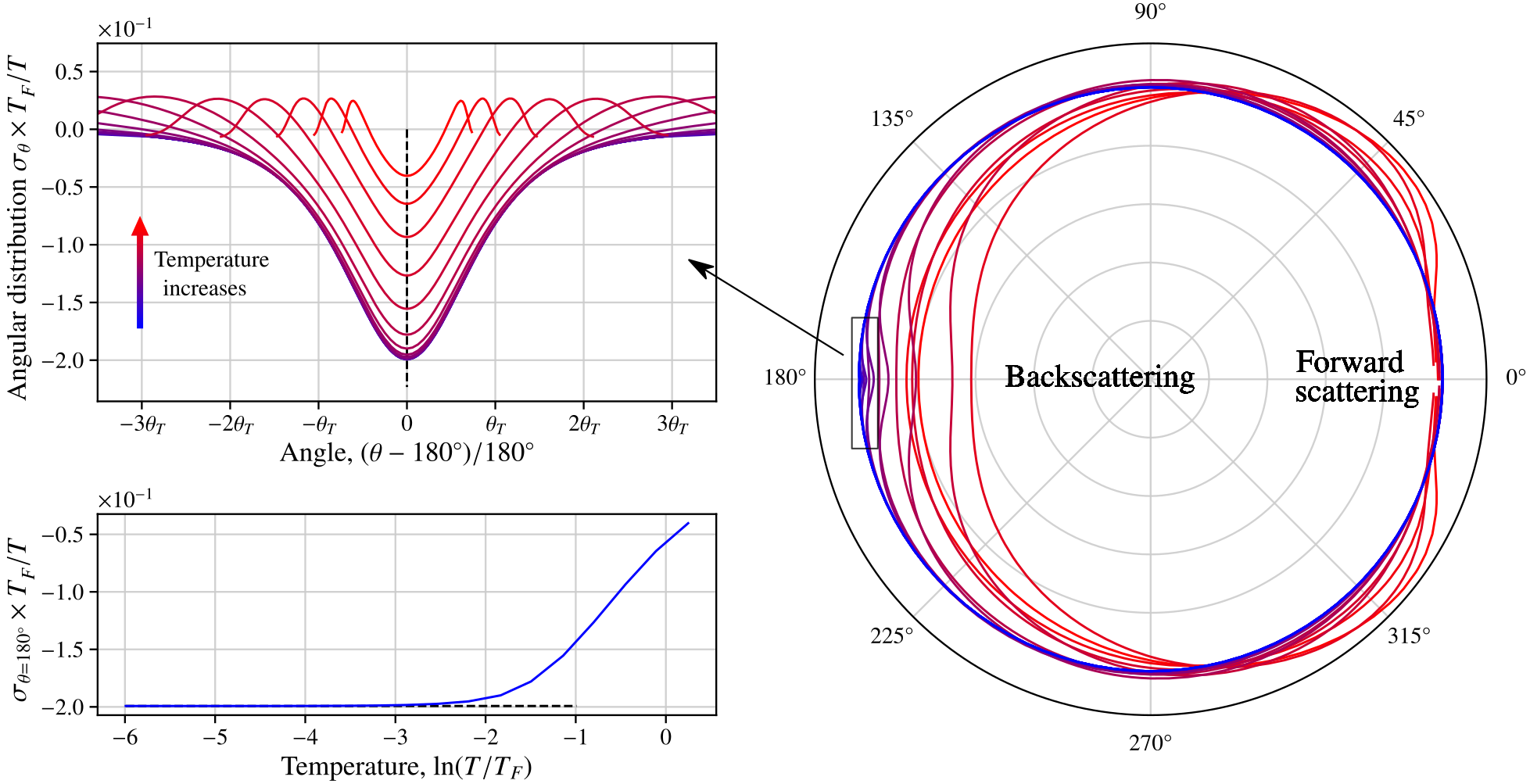}
	\caption{Angular distribution $\sigma(\theta)$ of scattered particles for different temperatures calculated for a two-body interaction $V$ with the angular dependence that suppresses
forward scattering. The temperatures used in this plot are
		$T/T_F = 0.0025, 0.005, 0.01, 0.02, 0.04, 0.08, 0.16, 0.32, 0.64, 1.28$. The value of the coefficient used in \eqref{eq:w_no_forw} is $a = 1/2$.}
	\label{fig:3plots_no_forw}
\end{figure}

\begin{figure}[h]
	\centering
	\includegraphics[width=0.45\textwidth]{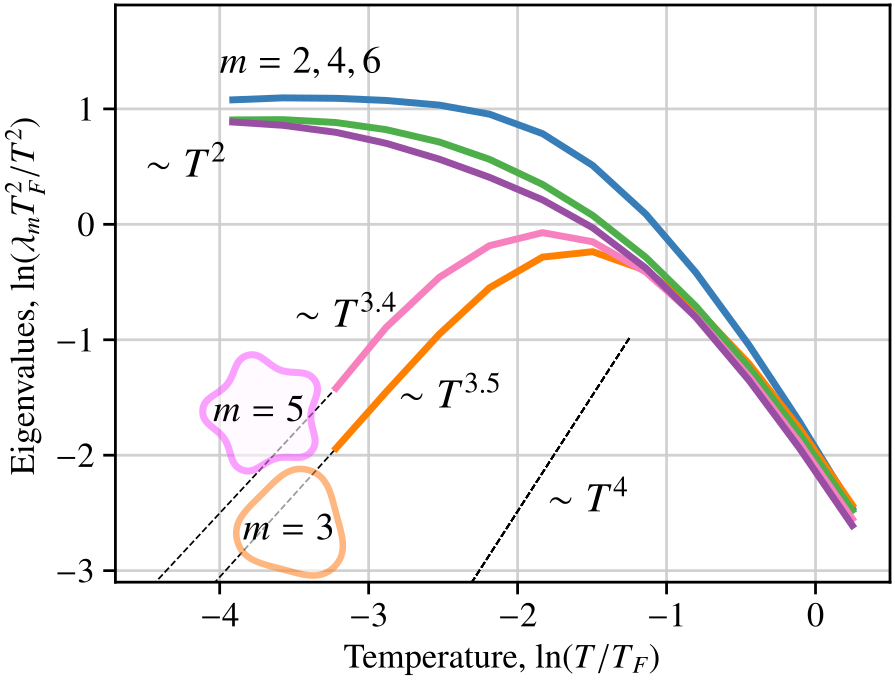}
	\caption{Eigenvalues for different harmonics as a function of temperature. Even $m$ eigenvalues show the $T^2$ scaling with temperature already at $T = 0.16 T_F$. Odd eigenvalues start to diverge from $T^2$ scaling to faster scaling regimes at temperatures lower than $T = 0.32 T_F$. The back-scattering shows $T^2$ dependence in the intensity $\sigma$ and $T$ dependence of the width of the scattering peak.}
	\label{fig:eigs_no_forw}
\end{figure}


\subsection{Separating contributions of the backscattering and forward scattering processes}

In this section we describe the method of separation of the backscattering from more noisy forward scattering and show that the details of forward scattering has no significant impact on the backscattering. 
We use this method to show that the potential numerical problems of our approach described above do not affect the qualitative features of the back-scattering and the eigenvalue hierarchy of odd and even harmonics. 
The initial assumption about the absolute value of the scattering matrix element in $|V| = \text{const}$.
Due to computational problems with precision of the forward scattering, we wish to study a modified potential that would prevent particles from scattering in a forward direction while keeping the backscattering effect intact. To accomplish this we choose a new 2-particle scattering potential that would satisfy $|V| \approx 0$ in the vicinity of $\vec{p}_1 \approx \vec{p}_1'$ and $\vec{p}_1 \approx \vec{p}_2'$ points. 
An example of a function that approaches zero when some $\vec{p}_1 \approx \vec{p}_1'$ is 
\begin{equation}
g(\vec{p}_1 - \vec{p}_1') = \left(1 - \exp \left[ - (\vec{p}_1 - \vec{p}_1')^2 / a^2 \right] \right).
\end{equation}

To study the dependence of the function above on the cutoff parameter $a$ we consider 4 values of $a = 1, 1/2, 1/4, 1/8$. To make the scattering matrix element to be explicitly symmetric under the time reversal symmetry and particle permutations, we construct it in a following way:
\begin{equation}
w(\vec{p}_1, \vec{p}_2, \vec{p}_1', \vec{p}_2') \sim 
|V|^2 g(\vec p_1 - \vec p_{1'}) \, 
g(\vec p_1 - \vec p_{2'}) 
g(\vec p_2 - \vec p_{1'}) \, 
g(\vec p_2 - \vec p_{2'})
\label{eq:w_no_forw}
\end{equation}
instead of $w(\vec{p}_1, \vec{p}_2, \vec{p}_1', \vec{p}_2') \sim |V|^2$.
We use the properly normalized scattering probability element from \eqref{eq:w_no_forw} with other parameters including temperature, integration mesh, and numerical integration precision being the same. This calculation repeats all the steps of the main calculation, but takes the transition element to approach zero in the case of forward scattering, which effectively turns it off and allows to omit the singularity during numerical integration of 
Eqs. \eqref{eq:I3} and \eqref{eq:I4}.

The plot of angle-resolved crosssection $\sigma(\theta)$ analogous to Fig. \ref{fig:3plots} 
of the main text is shown in Fig. \ref{fig:3plots_no_forw}. The backscattering with separated forward scattering in Fig. \ref{fig:3plots_no_forw} resembles the same qualitative properties as the backscattering in Fig. \ref{fig:3plots} 
of the main text while showing better numerical results and more abrupt regime change. The distribution enters the low-temperature scaling regime faster: the angular distribution already reaches $T^2$ scaling in the amplitude at the temperature $T = 0.32 \, T_F$, which can be observed at Fig. \ref{fig:3plots_no_forw}.

The deviation in behavior of the eigenvalues of odd $m$ harmonics from the behavior of eigenvalues of even $m$ harmonics shows up at the temperatures lower then $T = 0.32 \, T_F$, its behavior can be seen in Fig. \ref{fig:eigs_no_forw}. Besides the absence of the forward scattering, such calculation produced result analogous to the main results, which means that the forward scattering plays little role in creating the hierarchy of the even-$m$ and odd-$m$ eigenvalues.

\end{widetext}
\end{document}